# Enhancing Patent Retrieval Using Automated Patent Summarization*

Eleni Kamateri[1,*,†], Renukswamy Chikkamath[2,†], Michail Salampasis[1,†], Linda Andersson[3], and Markus Endres[2]

[1] *Department of Information and Electronic Engineering, International Hellenic University, Sindos, 57400, Thessaloniki, Greece*

[2] *Hochschule München, Loth str. 34, 80335 München, Germany*

[3] *Artificial Researcher - IT GmbH, Sweden*

**Abstract**

Effective query formulation is a key challenge in long-document Information Retrieval (IR). This challenge is particularly acute in domain-specific contexts like patent retrieval, where documents are lengthy, linguistically complex, and encompass multiple interrelated technical topics. In this work, we present the application of recent extractive and abstractive summarization methods for generating concise, purpose-specific summaries of patent documents. We further assess the utility of these automatically generated summaries as surrogate queries across three benchmark patent datasets and compare their retrieval performance against conventional approaches that use entire patent sections. Experimental results show that summarization-based queries significantly improve prior-art retrieval effectiveness, highlighting their potential as an efficient alternative to traditional query formulation techniques.

**Keywords**

patent retrieval, query formulation, patent summarization, big bird, summary segment

## 1 Introduction

Drafting a representative abstract that accurately summarizes the core concepts of an invention is av important step in the patenting process. A well-crafted abstract conveys the core concepts of an invention, therefore it enhances both the readability and discoverability of a patent throughout its lifecycle. For instance, integrating summaries into search snippets could reduce examiner search time, or help an inventor to quickly grasp prior art. A good summary may also be a great assistance for a patent professional to evaluate the technical or legal scope of a patent. Furthermore, a good summary retaining technical details and key claims could be used for downstream task such as patent prior-art and classification.

However, many human-authored patent abstracts do not summarize the invention effectively. This shortcoming may arise from various factors, such as the urgency to submit the application, regulatory constraints on abstract length, limited attention by inventors, and, last but not least, the intentional vagueness often employed to avoid narrowing the scope of legal protection and reduce discoverability in prior-art searches. Consequently, relying directly on the patent abstract for producing a search query for patent retrieval tasks is often ineffective.

As a result, patent abstracts are often supplemented by human-selected keywords. Alternatively, content is extracted from other sections of the patent, such as the description or claims, to produce queries that will enhance retrieval performance [1-3]. To address this need, various methods have been developed to automatically generate search queries from patent applications employing simple intuitive heuristics (e.g., the first X words), statistical techniques, language modeling and other methods. Some of these methods enhance abstracts by directly incorporating content from the description and claims sections [2]. Other methods aim to identify the most discriminative terms across different sections by comparing term statistics within a given patent to those of a broader corpus, often leveraging language modelling estimation techniques [4]. Query enrichment may also involve query terms extracted from patent citations or classification codes [5].

When using LLMs, there are two basic approaches for handling large documents -such as patents- for text retrieval and other text-related tasks. The first is document chunking to overcome token limits of LLMs. The second is document summarization which aims to preserve the semantic flow and concepts of a long document leading to a reduced text representation for efficient processing. AI-generated summaries are increasingly adopted across domains for their conciseness and informativeness. In the patent domain, however, their use has been primarily explored in classification tasks [6], where recent studies show that classifiers trained on generated summaries outperform those trained on original abstracts. Beyond classification, there is a growing need for purpose-specific summaries tailored to other downstream tasks. These summaries can take different forms depending on what the task requires (e.g., abstract, extended summary, or first claim), summarization approach (e.g., abstractive or extractive), input source (e.g., description, claims, or brief description) and output length (e.g., 50 to 300 words). One promising application is patent retrieval task, where high-quality and rightly sized summaries can serve as search queries that contain meaningful context but not so large that they overwhelm the model.

Building on these considerations, this study presents a pipeline for automated patent summarization. Furthermore, it examines the effectiveness of these automatically generated summaries as search queries for prior-art search, a critical task in the global patent operation. We employ state-of-the-art language models and semantically rich patent documents segments to generate both extractive and abstractive summaries, with the main goal to improve retrieval performance. We assess multiple input source combinations and summarization approaches across three patent datasets, determining which configurations produce the most informative summaries for retrieval tasks. Our results show that AI-generated summaries, when used as queries, consistently outperform other traditional strategies that rely on full patent sections.

The remainder of the paper is organized as follows: Sections 2 and 3 detail the semantically important segments found in patent description and present summarization techniques. Section 4 outlines the methodology adopted in our study. In Section 5 we present the experimental results, while in Section 6 we discuss the findings and conclude the paper.

## 2 Patent Description for Search

Among patent sections in a patent document, the description is consistently identified as the most informative and valuable source for query generation [2]. As the longest component, it offers a detailed elaboration of the proposed invention, often extending to several thousand words [7]. However, the absence of a standardized structure or mandated format for organizing patent description complicates automatic processing. This challenge is partially mitigated by the widespread use of conventional headings, such as *Background*, *Summary of the Invention*, *Brief Description of the Drawings,* and *Detailed Description of the Invention*. These headings are commonly adopted by patent applicants to organize the description into semantically coherent segments.

Notably, the combination of the *Background* and *Summary of the Invention,* which are collectively referred to as the "background summary", has proven to be the most effective source for extracting high-value query terms from U.S. patent documents [8]. The practical importance of these segments led to increased interest in their standardization. In response, the USPTO has expanded its public API to provide direct access to key description sub-sections, like the *Background* and *Brief Description*. The *Brief Description*, labeled as "brief" in the USPTO API, spans from the beginning of the description to the end of the *Summary of the Invention* segment, effectively capturing what is traditionally known as the "background summary".

Similarly, the Harvard USPTO Patent Dataset (HUPD) [9] uses USPTO documents' semi-structure and simple heuristics to extract more meaningful segments like the *Summary of the Invention*. This segment offers a more comprehensive and informative description of the invention substantially augmenting the abstract's content while improving classification performance [10].

Although these segments enhance patent retrieval and classification, they appear inconsistently across patent documents. For instance, these or other corresponding headings often do not exist in European patents. This structural inconsistency underscores further the motivation for automated methods capable of generating summary-like content (resembling the USPTO's *Summary of the Invention)*, particularly in documents that lack clearly defined segments in the description. To that end, summarization techniques play a key role in bridging this gap by producing coherent, informative summaries.

## 3 Patent summarization

Extractive and abstractive summarization represent two distinct approaches to generate summaries from text. Extractive summarization uses statistical or neural models to select the most relevant sentences directly from the source text, preserving the original text. In contrast, abstractive summarization uses transformer models to generate a condensed version of the content by rephrasing or synthesizing information using natural language generation techniques, producing more coherent and readable summaries. Finally, hybrid approaches may be used to combine accuracy (selecting existing text using extractive methods) as well as fluency by enhancing the previously extracted text with abstractive methods.

Typically, extractive models work by generating sentence-level embeddings, clustering these embeddings, and selecting the sentences nearest to the cluster centroids as the most representative. State-of-the-art models like BERT [11] and its variant SBERT [12] are widely used for this purpose. These models effectively extract the most informative sentences without altering them, maintaining the original phrasing and structure of the source document.

Abstractive summarization models, on the other hand, follow an encoder-decoder architecture: they encode the input text, generate a summary through a decoding process, and produce a fluent, often restructured, output [13, 14]. Notable models in this category include PEGASUS [15], T5 [16], and BART [17], which have demonstrated strong performance on long-document summarization tasks. Although GPT-based models [18] also demonstrate strong performance in abstractive summarization, their closed-source nature and token-based pricing present practical limitations for large-scale use.

In the patent domain, recent research has primarily focused on generating summaries from the description and/or claims sections using Large Language Models (LLMs) [9, 19]. We refer to these outputs as automated summaries, to distinguish them from the human-authored "*Summary of the Invention*" segments found within the description (hereafter referred to as summary segments). A number of previous studies have fine-tuned pre-trained language models on patent datasets. For instance, one early work [20] trained Seq2Seq [21], PointGenerator [22], and SentRewriting [23] models on the BIGPATENT dataset, using the description as input and the abstract as the target output. BigBird-Pegasus, a long-sequence transformer model, was later fine-tuned on BIGPATENT for improved summarization of patent texts [24]. Similarly, the work in [9] adapted two versions of the T5 model for patent data, using either the claims or the description sections to generate abstracts. Another study reported in [25] investigated which patent sections are most informative for generating the first independent claim (also referred to as the first claim), using PEGASUS and PointGenerator models, concluding that the summary segment is the most suitable input source for this task. Fine tuning and adaptation are crucial because generic pre-trained LLMs fail on technical/legal precision.

By examining both the structural characteristics of patent documents and the current state of research on patent summarization, we have identified several open issues that merit further investigation to improve the quality and utility of generated summaries, as follows:

i) Existing models are typically not trained on the full patent text. Due to token length constraints, typically limited to 512 or 1024 tokens in standard LLMs, such as T5, PEGASUS, and BART, and extended up to 4096 tokens or more in models like BigBird-Pegasus, the input text often needs to be truncated, segmented or adapted. This can hinder the model's ability to fully contextual capture understanding.

ii) Evaluation commonly relies on existing abstracts as ground-truth summaries, despite their frequent shortcomings in terms of clarity, completeness, and informativeness.

iii) A strong semantic alignment exists between the first claim and the summary segment. However, this relationship remains underexplored in current summarization approaches.

These limitations highlight the need for summarization strategies that are specifically tailored to the structure and practical use cases of patent documents. In particular, for retrieval tasks, such as prior-art search, where the generated summary serves as an effective retrieval query, it is essential to adopt summarization approaches that leverage high-value sections or combination of them to generate summaries that attain high retrieval performance. To operationalize this approach, our study follows a pipeline that first extracts key patent segments, then trains summarization models, and finally evaluates their effectiveness of the automated summaries in prior-art retrieval across multiple benchmark datasets.

# 4 Methodology

This study hypothesizes that generated summaries can improve the efficiency of prior-art search. To validate this hypothesis, we designed a five-stage experimental workflow, beginning with summary generation and ending with an evaluation of their effectiveness as retrieval queries. The performance of these summaries as queries is compared against sections, abstract, claims, and description, which are commonly used by patent professionals to formulate queries. In the following sections, we describe the data collections, we detail each stage of the methodology, and explain the evaluation process used to assess both the intrinsic quality of the generated summaries and their impact on retrieval performance.

## 4.1 Data collections

We utilize four patent datasets, each serving a distinct role. The first dataset, HUPD, allows the extraction of salient sections from patent documents. The second dataset, BIGPATENT, is employed for the intrinsic evaluation of the generated summaries, i.e. to measure the quality of the automated summaries compared to reference summaries. Finally, the next two datasets, CLEF-IP 2013 and USPTO, are used for extrinsic evaluation applying summaries for prior-art search.

**HUPD [9]:** The HUPD is a large-scale, structured corpus of English-language utility patent applications filed to the USPTO between 2004 and 2018. Each JSON-formatted entry contains rich metadata, including bibliographic details, classification codes, inventor information, and full text fields such as abstract, claims, background, summary, and description.

**BIGPATENT [20]:** The BIGPATENT dataset is a large-scale patent summarization benchmark comprising approximately 1.3 million U.S. patent documents. It pairs the description section with its corresponding abstract, serving as a ground truth for training and fine-tuning summarization models. For our intrinsic evaluation, we randomly sampled 1,000 patents from the available 67,072 of the test set.

**CLEF-IP [31]**: The CLEF-IP collection consists of patent documents sourced from the EPO and WIPO. The English topic set from the CLEF-IP 2013 campaign originally comprises 50 topics [26]. However, because the topics are based on patent claims rather than individual documents, there is no strict one-to-one mapping between topics and documents. In total, these topics correspond to 37 unique documents. Due to missing relevant documents for some topics in the indexed dataset, we further reduced the set to 24 English-language patents for our experiments. Each topic patent is associated with between 2 and 8 manually identified relevant documents, based on expert-curated citation links, making this dataset a reliable benchmark for evaluating prior-art retrieval performance.

**USPTO - Explainable AI for Patent Professionals [27] (referred to as USPTO):** This dataset was released as part of a Kaggle competition aimed at advancing explainable AI in the patent domain. Each topic patent is associated with a set of 50 most similar patents, identified using content similarity measures rather than citation-based relevance. From this dataset, we selected 3,343 topic patents in which semantically coherent segments were automatically detected. Unlike CLEF-IP, which relies on citation-based ground truth, this benchmark provides an opportunity to test retrieval performance under automated similarity-based relevance.

## 4.2 Patent part extraction

The first step in our pipeline focuses on identifying and extracting sections of each patent to serve as query sections or as input sources for summarization. Specifically, we use the description and claims sections, and when identifiable, include the brief description, summary and first claim.

To detect these segments, we utilized the HUPD dataset, which is currently the only resource providing annotated labels (i.e., tags) for the background and summary segments within the description section of US patent documents. Based on these annotations, we constructed a dictionary of relevant summary headings, which we then used as a reference to identify candidate headings in unannotated patents. For each heading labeled as a summary heading, the subsequent content was marked as a summary segment. Once the summary segment was identified, the brief description was also derived by selecting the text spanning from the beginning of the description section up to the end of the summary. Finally, the first claim was extracted using heuristic rules, specifically, by identifying the first claim that is not dependent on any previous claim.

## 4.3 Patent summarization

In this phase, we employ three summarization models, the BERT, SBERT and BigBird-Pegasus (from now on referred to as BigBird). BERT and SBERT are utilized for extractive summarization, focusing on identifying the most relevant sentences from the input text. BigBird, which has been pre-trained on the BIGPATENT dataset, serves as our primary abstractive summarization model, to handle long-form patent text effectively [24].

For BERT and SBERT, we used models' default configurations. Specifically, for SBERT, we employed the "paraphrase-MiniLM-L6-v2" model. For BigBird, we used a model fine-tuned on the BIGPATENT dataset. Two configuration variants of the BigBird model were explored: the default configuration, which generates relatively short summaries (typically between 50 and 100 words, depending on the input), and a modified version, where the model's generation parameters were adjusted (i.e., the length penalty and minimum/maximum length settings) to produce longer summaries ranging from 250 to 300 words

In this study, the BigBird model is further fine-tuned to generate summaries using the brief description and first claim as inputs, which are two sections identified as particularly informative within patent documents. The target output for this fine-tuning process is the summary segment. This exploration aims to set the foundation for future research on fine-tuning summarization models to replicate other valuable parts of patent text, such as the extended, author-crafted summary segments found within the description section.

**Table 1**
Fine-tuning Parameters

| Parameters | Value |
|---|---|
| max_source_length | 1024 |
| num_beams | 4 |
| length_penalty | 0.8 |
| no_repeat_ngram_size | 3 |
| max_target_length/min_target_length | 300/100 |
| learning_rate | 2e-5 |
| per_device_train_batch_size | 1 |
| gradient_accumulation_steps | 16 |
| num_train_epochs | 2 |

To achieve this, a new dataset, which is a subpart of the HUPD dataset, is specially created. Specifically, we extracted from the HUPD dataset 402,921 patents that have a distinct summary segment with a length between 150 and 250 words. Then, we extracted the brief description and first claim and selected those patents whose brief description and first claim together had a length between 700 and 800 words. This selection criterion allows to skip any adjustment steps of the input text during the training, such as truncation or chunking, which may negatively affect the model's interpretation. All these steps led to a dataset comprising 48,322 patents, which was finally used to fine-tune the BigBird model. An overview of the summarization and training parameters is shown in Table 1.

### 4.4 Patent retrieval

For the patent retrieval task, we use the FAISS vector database to store and retrieve semantic vectors. Both queries and patent documents are embedded using the GTE-large-en-v1.5 model [28], which has approximately 409 million parameters, a 1,024-dimensional embedding size, and supports input lengths up to 8,192 tokens. This enables the generation of rich embeddings that effectively represent full patents. The model achieved state-of-the-art performance on the Massive Text Embedding Benchmark (MTEB) within its size category, making it well-suited for our application. To reduce hardware complexity, we limit input length to 3,000 tokens, which is sufficient to capture both independent and dependent claims and is used as corpus embeddings. While alternative embedding models and strategies, such as using different sections (e.g., abstracts, descriptions) or specific segments (e.g., brief descriptions, summaries) as corpus representations offer promising avenues for exploration, we leave these investigations for future work.

Since our primary goal is to assess the impact of generated summaries on prior-art retrieval, we restrict the vector index to a subset of 200,000 patent documents drawn from our two prior-art datasets. Retrieval performance is then compared against simpler methods that use entire patent sections or extracted segments as queries. Given that our summarization pipeline is based on semantic models, and our aim is to isolate the contribution of the generated summaries from the retrieval technique itself, we focus exclusively on embedding-based retrieval method rather than keyword-based approaches. The integration of additional retrieval methods is left for future work.

### 4.5 Intrinsic and extrinsic evaluation

To evaluate the effectiveness of the generated summaries, we perform two types of evaluations: intrinsic and extrinsic.

For the intrinsic evaluation, the automatically generated summaries are compared against reference summaries, either the original patent abstract or the annotated summary segment (which has been proved to provide an extended and improved summary). This evaluation aims to determine how accurately the generated summaries captures the key content of the patents. For this purpose, we use ROUGE scores [29] to assess textual overlap and compute semantic similarity, calculated as the cosine similarity between embeddings produced by Google's BERT-for-Patent model [30], comparing the generated and reference summaries.

For the extrinsic evaluation, we measure the impact of using the generated summaries as queries in a prior-art search task. Specifically, we compare their retrieval performance against traditional query strategies using standard IR metrics, including Mean Average Precision (MAP), Precision, and Recall.

# 5 Results

## 5.1 Evaluation of summary generation in the BIGPATENT dataset

Table 2 presents Rouge-1, Rouge-L and semantic similarity scores between the generated summaries and the reference summaries, which consists of either the original abstract or the extracted summary segment. The summaries were generated using the pre-trained BigBird model, with inputs from the description section (averaging around of ~2,150 words) and the extracted brief description (averaging approximately ~650 words).

When abstract is taken as the reference summary, the fine-tuned BigBird model achieves similar evaluation scores with the pre-trained model, especially for brief description. This is an improved feature considering that a similar quality summary is produced with less content, as the brief description is a subpart of the description.

In terms of the summary reference, the fine-tuned BigBird model is the best performed, generating better quality and similarity summaries to extracted summary segment. This is natural since our model has been specially trained to target this specific segment, which is considered a good candidate to substitute the abstract.

Our experiments indicate that fine-tuning summarization models to generate summaries targeting key segments provides a strong foundation for adapting these models to replicate valuable components of patent text, such as the extended, author-crafted summary segments found within the description section.

## 5.2 Evaluation of prior-art retrieval in the CLEF-IP 2013 dataset

In the CLEF-IP 2013, we followed the TREC-based guidelines provided by CLEF-IP [31], using TRecTools [32] to calculate Precision and Recall at various cut-off levels (@5, @10 and @30), as well as MAP@100.

Table 3 reports the retrieval results of conventional query strategies, where entire patent sections commonly used by professionals are employed verbatim as queries. We observe that queries formulated using the claims text achieved the best retrieval results. This outcome is largely attributed to the fact that the claims text was also used for generating the corpus embeddings, thereby ensuring a higher degree of semantic alignment between the query and the indexed documents. While extracting patent segments for query formulation could offer valuable insights, it was not feasible to implement this approach effectively, since the description sub-sections were not consistently identifiable across all 24 topics.

Table 4 presents the retrieval results when using automated summaries as query inputs. These summaries were generated using various summarization methods and different patent sections as input sources. The results demonstrate that queries formulated with these automated summaries consistently outperform those based on standard patent sections. Overall, the default summaries generated by the BigBird model although outperform the standard patent sections, they were found to be insufficient in capturing the breadth of important patent content compared to longer summaries generated by the adjusted BigBird model, which achieve best retrieval performance.

**Table 2**

ROUGE and semantic similarity scores in BIGPATENT dataset. The best results per reference summary are in bold.

| Reference | Method | Input | Avg. #words | Rouge-1 | Rouge-L | Semantic Similarity |
|---|---|---|---|---|---|---|
| Abstract | BigBird* | Description | **118** | **0.51** | **0.42** | **0.81** |
|  | FT BigBird | Description | 157 | 0.40 | 0.29 | 0.78 |
|  | BigBird* | Brief descr. | 50 | 0.45 | 0.37 | 0.78 |
|  | FT BigBird | Brief descr. | 122 | 0.47 | 0.35 | 0.81 |
| Summary | BigBird* | Description | 118 | 0.50 | 0.48 | 0.66 |
|  | FT BigBird | Description | 157 | 0.50 | 0.46 | 0.68 |
|  | BigBird* | Brief descr. | 50 | 0.28 | 0.25 | 0.66 |
|  | FT BigBird | Brief descr. | **122** | **0.56** | **0.53** | **0.74** |

*: Pre-trained BigBird (default), FT: Fine-tuned BigBird

**Table 3**
Retrieval results in CLEF-IP dataset using patent sections as queries. The best results are in bold.

| Section | Avg. #words | MAP@100 | Precision | | | Recall | | |
|---|---|---|---|---|---|---|---|---|
| | | | @5 | @10 | @30 | @5 | @10 | @30 |
| Abstract | 109 | 26.31% | 20.00% | 14.58% | 7.08% | 26.46% | 35.80% | 50.27% |
| Claims | 982 | 27.72% | **23.33%** | **15.83%** | **6.81%** | **28.30%** | **36.40%** | **47.17%** |
| Description | 6,962 | 23.89% | 20.00% | 12.50% | 5.42% | 22.47% | 28.12% | 35.87% |

**Table 4**
Retrieval results in CLEF-IP dataset using generated summaries as queries. The best results per section are in bold.

| Summary Source | Method | Avg. #words | MAP@100 | Precision | | | Recall | | |
|---|---|---|---|---|---|---|---|---|---|
| | | | | @5 | @10 | @30 | @5 | @10 | @30 |
| Claims | BERT | 155 | 31.38% | 25.00% | 15.00% | 6.81% | 29.38% | 35.07% | 46.92% |
| | SBERT | 157 | 29.43% | 23.33% | 15.00% | 6.53% | 26.60% | 33.38% | 43.91% |
| | BigBird* | 62 | 31.60% | 24.17% | 15.83% | 7.08% | 28.93% | 37.57% | 49.94% |
| | BigBird** | 224 | **35.40%** | **27.50%** | **18.33%** | **7.78%** | **32.12%** | **42.30%** | **53.07%** |
| Description | BERT | 1121 | 30.59% | 25.00% | 15.83% | 6.94% | 28.56% | 35.95% | 49.63% |
| | SBERT | 1,276 | 30.89% | 25.00% | 15.42% | 7.08% | 28.93% | 34.84% | 51.33% |
| | BigBird* | 123 | 28.00% | 22.50% | 14.17% | 6.81% | 26.85% | 32.16% | 48.31% |
| | BigBird** | 239 | **32.69%** | **26.67%** | **16.67%** | 6.94% | **30.73%** | **38.72%** | 48.13% |

*: Pre-trained BigBird (default), **: Adjusted pre-trained BigBird, FT: Fine-tuned BigBird

Then comes the SBERT-based summaries, particularly when the description text was used as input, or the default BigBird when using the claims text. Although the BERT- and SBERT-based summaries achieved good retrieval scores, especially when produced from the description, they often retained much of the original text without adequately condensing it, which is crucial for overcoming the token limitations of LLMs in text retrieval tasks.

### 5.3 Evaluation of prior-art in the USPTO dataset

In the USPTO dataset, we followed the USPTO Kaggle competition guidelines, computing MAP@50 as the primary metric. For Precision and Recall, we applied the same cut-off levels used in the CLEF-IP 2013 evaluation (@5, @10 and @30) to ensure consistency and comparability across datasets.

Table 5 presents retrieval results based on traditional patent sections commonly used as queries, such as the abstract, claims, and description. It also includes results from using high-value segments, such as the summary segment, brief description, and first claim, individually and in combination. Interestingly, the queries formulated by these high-value segments consistently outperform those based on conventional patent sections. This underscores the importance of targeted content selection in enhancing retrieval effectiveness.

Table 6, on the other hand, reports retrieval performance when automated summaries are used as queries. Then, we observe that depending on the summarization method and input, automated summaries can significantly outperform their respective original patent sections.

In particular, the adjusted BigBird model, which generates longer summaries of approximately 250-300 words compared to the default version, outperforms the default BigBird model in retrieval performance. Furthermore, it achieves results that are comparable to, or slightly better than, those obtained using the simpler query formulation techniques outlined in Table 5. Notably, this approach demonstrates strong efficiency, as it achieves similar retrieval performance using generated summaries that are substantially more concise than the original patent sections.

Regarding extractive models, queries generated using SBERT summaries based on the description text achieved the highest retrieval scores across all metrics. In contrast, queries generated from the claims or brief description text performed worse than those using the original texts. Interestingly, despite their strong retrieval performance, SBERT-generated summaries, averaging 807 words. This contrast highlights the importance of aligning summary generation with its intended purpose, whether to enhance readability and support human assessment, or to optimize performance in downstream tasks such as prior-art retrieval and classification.

**Table 5**
Retrieval results in USPTO dataset using patent sections and extracted patent segments as queries. The best results are in bold.

| Patent Section/Extracted Segment | Avg. #word | MAP@50 | Precision | | | Recall | | |
|---|---|---|---|---|---|---|---|---|
| | | | @5 | @10 | @30 | @5 | @10 | @30 |
| Abstract | 139 | 16.60% | 51.18% | 44.77% | 32.84% | 5.12% | 8.95% | 19.70% |
| Claims | 832 | 20.17% | 60.60% | 52.82% | 37.88% | 6.06% | 10.56% | 22.73% |
| Description | 4,200 | 21.19% | 62.54% | 55.16% | 40.06% | 6.25% | 11.03% | 24.04% |
| Brief desc | 1,096 | 22.40% | 63.94% | 56.42% | 41.63% | 6.39% | 11.28% | 24.98% |
| Summary | 524 | 21.08% | 60.34% | 53.52% | 39.52% | 6.03% | 10.70% | 23.71% |
| Summary and first claim | 695 | 22.50% | 63.73% | 56.62% | 41.34% | 6.37% | 11.32% | 24.80% |
| Brief desc and first claim | 1,267 | 22.64% | **64.18%** | **56.91%** | **41.89%** | **6.42%** | **11.38%** | **25.13%** |

**Table 6**
Retrieval results in USPTO dataset using automated summaries as queries. The best results per section are in bold.

| Summary Source | Method | Avg. #words | MAP@50 | Precision | | | Recall | | |
|---|---|---|---|---|---|---|---|---|
| | | | | @5 | @10 | @30 | @5 | @10 | @30 |
| Claims | BERT | 100 | 12.98% | 41.33% | 36.09% | 26.32% | 4.13% | 7.22% | 15.79% |
| | SBERT | 104 | 13.36% | 42.04% | 36.90% | 26.81% | 4.20% | 7.38% | 16.08% |
| | BigBird* | 74 | 15.28% | 47.65% | 42.31% | 31.22% | 4.76% | 8.46% | 18.73% |
| | BigBird** | 227 | **19.72%** | **57.73%** | **51.10%** | **37.65%** | **5.77%** | **10.22%** | **22.59%** |
| Description | BERT | 695 | 23.40% | 64.36% | 57.71% | 42.91% | 6.44% | 11.54% | 25.74% |
| | SBERT | 807 | **23.95%** | **65.84%** | **58.78%** | **43.68%** | **6.58%** | **11.76%** | **26.21%** |
| | BigBird* | 79 | 15.50% | 49.57% | 43.57% | 31.87% | 4.96% | 8.71% | 19.12% |
| | BigBird** | 242 | 21.89% | 62.60% | 55.39% | 40.82% | 6.26% | 11.08% | 24.49% |
| Brief desc | BERT | 205 | 19.95% | 57.74% | 51.25% | 38.06% | 5.77% | 10.25% | 22.84% |
| | SBERT | 225 | **20.66%** | **59.41%** | **52.57%** | **39.01%** | **5.94%** | **10.51%** | **23.41%** |
| | BigBird* | 64 | 15.71% | 49.48% | 43.71% | 31.98% | 4.95% | 8.74% | 19.19% |
| | BigBird** | 231 | 21.18% | 60.83% | 54.09% | 39.88% | 6.08% | 10.82% | 23.93% |

*: Pre-trained BigBird (default), **: Adjusted pre-trained BigBird, FT: Fine-tuned BigBird

## 6 Discussion and conclusion

Overall, the findings of this study are promising, demonstrating that patent retrieval benefits from using targeted patent segments (when detectable) and automated summaries as queries, compared to relying solely on traditional sections typically employed by patent professionals, such as abstract, descriptions or claims. Across both prior-art datasets, CLEF-IP and USPTO, automated summaries consistently outperformed conventional query inputs. Among the summarization methods and input configurations evaluated, the adjusted BigBird model using claims as input and the SBERT model applied to the description section emerged as the most effective abstractive and extractive approaches, respectively, yielding the highest retrieval performance across both datasets.

Moreover, our initial experimental results support the hypothesis that summarization models can be further adapted to produce comprehensive and contextually relevant summaries, although confirming this required extensive validation. This approach presents a promising direction for future advancements in patent summarization.

This work was initially motivated by our participation in the European Patent Office's (EPO) CodeFest 2024 competition [33], where it was selected as one of the top six finalists. Building on this foundation, we aim to further advance our research on patent segmentation and summarization techniques by evaluating their impact on patent retrieval performance across additional prior art test collections. Additionally, exploring alternative corpus representations and integrating additional retrieval methods will be key areas of focus in future work.

## Acknowledgements

This research work was supported by the Hellenic Foundation for Research and Innovation (HFRI) under the HFRI PhD Fellowship grant (Fellowship Number: 10695).

## A. Online Resources

The link to the GitHub repository will be made available and open to the public after the publication of the paper.